\documentclass{aa}

\usepackage{amsmath, amssymb}
\usepackage{commath}
\usepackage{nicefrac}
\usepackage{hyperref}
\usepackage{mathtools}
\usepackage{blindtext}

\usepackage{booktabs}

\usepackage[dvipsnames]{xcolor}
\usepackage{soul, color}
\usepackage[varg]{txfonts}

\usepackage{natbib}
\bibpunct{(}{)}{;}{a}{}{,} 

\edef\TeXjobname{\jobname} 
\edef\jobname{\detokenize{aa58167-25}}

\begin{document}

\title{Exploring rotational properties and the YORP effect in asteroid families}

\author{G. Bertinelli\inst{\ref{inst1}} 
\and W.-H. Zhou\inst{\ref{inst2}}\fnmsep\thanks{Corresponding author} 
\and P. Tanga\inst{\ref{inst3}}}

\institute{Department of Physics and Astronomy, University of Padova, Padova, Italy\label{inst1}\\
\email{gabriele.bertinelli@studenti.unipd.it}
\and JSPS International Research Fellow, Department of Earth and Planetary Science, University of Tokyo, Tokyo, Japan\label{inst2}\\
\email{wenhan.zhou@oca.eu}
\and Université Côte d'Azur, Observatoire de la Côte d'Azur, CNRS, Laboratoire Lagrange, Bd de l'Observatoire, CS 34229, 06304 Nice Cedex 4, France\label{inst3}}




\date{Received 18 November 2025 / Accepted 8 December 2025}

\abstract{The long-term dynamical evolution of asteroid families is governed by the interplay between orbital and rotational evolution driven by thermal forces and collision.}{We aim to observationally trace the rotational evolution of main-belt asteroid families over Gyr timescales.}
{{We analyzed rotational properties of 8739 asteroids with spin period measurements and 3794 asteroids} with obliquity determinations across 28 asteroid families spanning ages from 14~Myrs to 3~Gyrs. We introduced a dimensionless timescale that normalizes each asteroid's family age by its classical YORP timescale, enabling direct comparison of rotational states across different evolutionary stages. We examined two key observables: the fraction of slow rotators (periods greater than or equal to 30 hours) and the polarization fraction (the degree to which asteroid spin poles align correctly with their position in the family's V-shape distribution according to the Yarkovsky theory). Evolution of both quantities were fitted to identify characteristic transition timescales.}
{We discovered that the slow-rotator fraction increases steeply with $t$ and saturates at $f_{\rm slow} \simeq 0.25$ around a breakpoint $t_{\rm bp} \simeq 20$. This implies a stochastic YORP timescale $\tau_{\rm YORP,stoc} \simeq 10\,\tau_{\rm YORP}$ by comparison with rotational evolution models that include tumbling and weakened YORP torques. The polarization fraction reaches a maximum of $\simeq 0.8$ at $t \simeq 16$ and then decays toward the random limit $f_{\rm pol} \rightarrow 0.5$ for $t \gtrsim 20$, indicating an increasing dominance of collisional spin reorientation over time.}
{The rotation properties within different asteroid families offer crucial clues to rotation evolution and can serve as a new dimension for age estimation of asteroid families with more data in the LSST era.
}
\maketitle

\nolinenumbers

\section{Introduction}\label{sec:introduction}

The modern asteroid belt is primarily composed of collisional fragments of planetesimals. Groups of fragments that share a common parent body are known as asteroid families \citep{Hirayama1918}. The long-term dynamical evolution of asteroid families in the Solar System is significantly influenced by non-gravitational forces, such as the Yarkovsky effect, which arises from the anisotropic emission of thermal photons from the surface of a rotating body, resulting in an acceleration that secularly modifies the body's semi-major axis.

The Yarkovsky-induced orbital drift of asteroids in the main belt is {$\sim 10^{-5} -10^{-4} ~\rm au~Myr^{-1}$ for 1~km objects and scales as $D^{-1}$ with $D$ being the asteroid diameter \citep{Bottke2006, spoto_asteroid_2015}}, leading to the orbital dispersion of the asteroid family. The predictable nature of the Yarkovsky effect has made it an invaluable tool for estimating the ages of asteroid families. Over time, this size-dependent dispersion creates a characteristic V-shape in a plot of a family's semi-major axis versus the inverse diameter of its members. The analysis of this V-shape, combined with an independent evaluation of the Yarkovsky drift rate, allows for an estimation of the time elapsed since the formation of the family {\citep[e.g.][]{Vokrouhlicky2006, walsh_introducing_2013, MILANI201446, delbo_primordial_2017}}.

However, the evolution is complicated by the dependence of the Yarkovsky effect on several parameters that are difficult to determine, e.g., thermal inertia and bulk density. Assuming similar composition and surface properties across a given family, the remaining dependencies are spin rate and obliquity. The rotation of asteroids is mainly affected by collisions and a thermal torque known as YORP (Yarkovsky-O'Keefe-Radzievskii-Paddack) that is induced by the scattering and re-emission of solar radiation from an asteroid's irregular surface \citep{paddack_1969, rubincam_radiative_2000}. The YORP effect applies a torque that, over long timescales, can significantly modify its spin vector, altering both the spin period and the spin-axis obliquity. Collisions dominate large and slowly rotating asteroids, whereas YORP is more effective for small and fast rotators \citep{Pravec2008}.

In recent years, data from the Gaia mission \citep{gaia_dr3_collab} have led to the computation of a vast number of new asteroid spin vectors \citep{durech_spins_2023, cellino_spins_2024}. This wealth of data has revealed new and unexpected trends. For example, \citet{durech_spins_2023, cellino_spins_2024, Vavilov2025} analyzed the spin-period distribution and confirmed an excess of slow rotators that are separated by a `valley' from the population of fast rotators in the period-diameter diagram. This excess of slow rotators is explained by \citet{wenhan_slowrot_25} as the result of the gathering of the slowly tumbling asteroids (i.e., asteroids in non-principal axis rotation) that suffer from a weaker YORP torque.

This work aims to investigate the rotational evolution of asteroids among asteroid families from the perspective of observation. We introduce a new parameterization built on general features of the Yarkovsky drift that are common to different families. As described in Sect. \ref{sec:methods}, by exploiting the estimated family ages, we introduce a normalized timescale that allows us to build a common framework of interpretation. Our results (Sect. \ref{sec:implications}) show that the current dataset of rotational properties for the members of 28 families provides useful new constraints to the evolution timescales of the asteroid rotations. In Sect. \ref{sec:conclusions} we summarize our findings.

\begin{figure}
    \centering
    \includegraphics[width=0.49\textwidth]{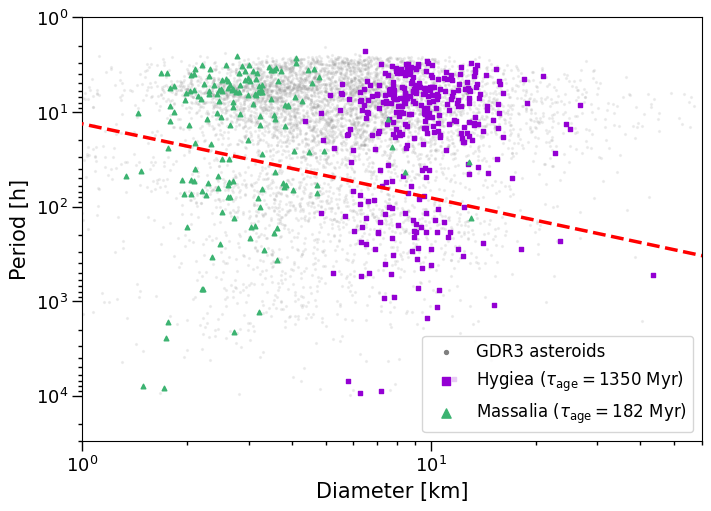}
    \caption{In gray, observational data from Gaia showing the period-diameter distribution for asteroids. As purple squares the distribution of the Hygeia family, and as green triangles the distribution of the Massalia family. The red dashed line is the fitted line that identifies the gap \citep{wenhan_slowrot_25}. Spin period values were calculated by \citet{durech_spins_2023} and \citet{cellino_spins_2024}. The ages in the legend are the weighted average ages computed from \citet{spoto_asteroid_2015}.}
    \label{fig:slowrot}
\end{figure} 

\section{Methods}\label{sec:methods}
We selected 28 asteroid families with ages ranging from 14.34 Myr (Nele) to 3024.13 Myr (Themis) \citep{spoto_asteroid_2015}. The families and their properties are listed in Table~\ref{tab:families}. The membership of each asteroid to a family and the proper elements were retrieved from the Asteroid-Dynamic Site (AstDyS, \citet{MILANI201446}). {We are aware that the estimation of family ages could be affected by methodological limitations and uncertainties in the Yarkovsky calibration. These problems will be addressed in Sect.~\ref{sec:errors}, along with assessing the impact of errors, when estimating the diameter, on the final result.}\\
{Diameter $D$, and albedo $p_V$ were extracted from the WISE/NEOWISE survey \citep{masiero_main_2011}}. For objects observed multiple times by the survey, we took the weighted average of their measurements {using as weights the errors on $D$ and $p_V$. Since the absolute magnitude $H$ does not have an error associated with the measure, we just calculated the mean}.\\
In order to increase the statistics, for objects that do not have a measurement for the diameter, we calculated it using 
\begin{equation}
D = \frac{1329}{\sqrt{\hat{p}_V}} \cdot 10^{-0.2 H}\,\,[{\rm km}]\,\, ,
\end{equation}
where $\hat{p}_V$ is the weighted average albedo of the family and {the absolute magnitude $H$ was retrieved from the Asteroid Family Portal \citep{AFP_2022}. $\hat{p}_V$ is calculated after having discarded possible outliers, using the interquantile range.
To calculate the error associated with $\hat{p}_V$, we performed a Monte Carlo bootstrap sampling of the errors associated with $p_V$ and took the median of this set. We repeated the operation for 5000 times. In the end we calculated the standard deviation of the set of the medians calculated above}.

Spin vectors were retrieved from \citet{cellino_spins_2024} and \citet{durech_spins_2023}. In addition, we also included data from the Asteroid Lightcurve Database (\href{https://minorplanet.info/php/lcdb.php}{LCDB}). From this last source, we selected only observations with a quality flag $Q\ge 3-$. For objects that have more than one suitable measurement, we selected the most recent one \citep{lcbd}. In the end, we obtained 8739 objects with a spin period and 3794 objects with obliquity estimation. {The complete dataset is reported in the Zenodo repository}\footnote{\href{https://doi.org/10.5281/zenodo.18105913}{https://doi.org/10.5281/zenodo.18105913}}.\\
The lower statistics regarding the obliquity can be attributed to the difficulty in reconstructing light curves. {The period-diameter distribution of our sample is presented in Fig.~\ref{fig:slowrot}, which further shows that this `valley' feature is present not only in the general asteroid population, but also within individual families}.
It is important to note that the dataset is affected by observational biases. For example, it is easier to determine the spin state of fast rotators with respect to slow rotators because the latter require a longer observation time to cover a full rotation. {Detecting super-fast rotators (SFRs, $P < 2.2$ h) can also be challenging, particularly in sky surveys, due to an insufficient observation cadence \citep{novakovi_sfr_2025}. However, in our dataset only two asteroids, namely 1999 RF226 and 2008 VN1, can be classified as SFRs}. In addition, the dataset is biased toward brighter asteroids, which are easier to observe.

\begin{figure}
    \centering
    \includegraphics[width=0.49\textwidth]{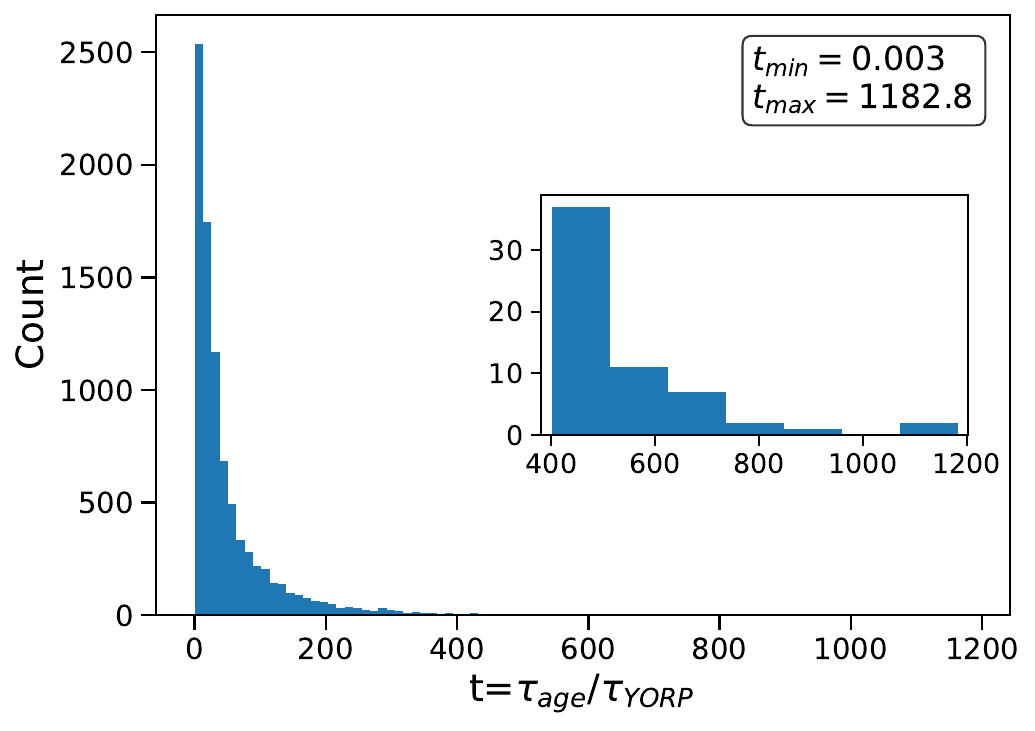}
    \caption{Distribution of calculated $t$ values for 8739 asteroids. The minimum $t$ is $t_{\rm min} \simeq 0.003$, while the maximum is $t_{\max} \simeq 1182.8$.}
    \label{fig:t_distr}
\end{figure}

We define a dimensionless time $t$ for each asteroid 
\begin{equation}
t = \frac{\tau_{\text{age}}}{\tau_{\text{YORP}}}\; ,
\label{eq:t}
\end{equation}
where $\tau_{\text{age}}$ is the age of the family. {\citet{spoto_asteroid_2015} estimated the ages for both sides of the V-shape, because families can be non-symmetrical in the $(a,1/D)$ space (e.g., in young families due to differences in the ejection velocities or in old families due to different impacts). For simplicity, we adopt the weighted average of the two age estimations, using as weights the errors associated with the measures. The differences are taken into account in the error bars and will be addressed in Sect.~\ref{sec:errors}}.\\ 
$\tau_{\text{YORP}}$ is the normal YORP timescale \citep{rubincam_radiative_2000, Vokrouhlicky2002, Golubov2019} 
\begin{equation}
    \tau_{\text{YORP}} \sim 1 \,\bigg(\frac{D}{1\, \text{km}}\bigg)^2\bigg(\frac{a_p}{2.5\, \text{AU}}\bigg)^2 \quad [\text{Myr}]\; ,
    \label{eq:yorp_time_scale}
\end{equation} 
where $D$ is the diameter of the asteroid and $a_p$ is the proper semi-major axis. The YORP timescale is defined as the time it would take for an asteroid to double its rotation rate under the YORP torque \citep{rubincam_radiative_2000}. The distribution of calculated $t$ is shown in Fig.~\ref{fig:t_distr}

\subsection{Slow rotators fraction}\label{sec:slow_frac}

The excess of slow rotators was discovered in the last century \citep[e.g.][]{Harris1979, Pravec2000, durech_spins_2023}. \citet{wenhan_slowrot_25} shows that they could be in a tumbling state caused by YORP de-spin. As a result, this excess of slow rotators is expected to emerge on $\sim 2$ YORP timescales. To examine this condition, we investigate the fraction of slow rotators and its evolution among asteroid families as their ages are known, assuming that the initial spin state is set immediately after the family forming event.

\begin{figure}
    \centering
    \includegraphics[width=0.49\textwidth]{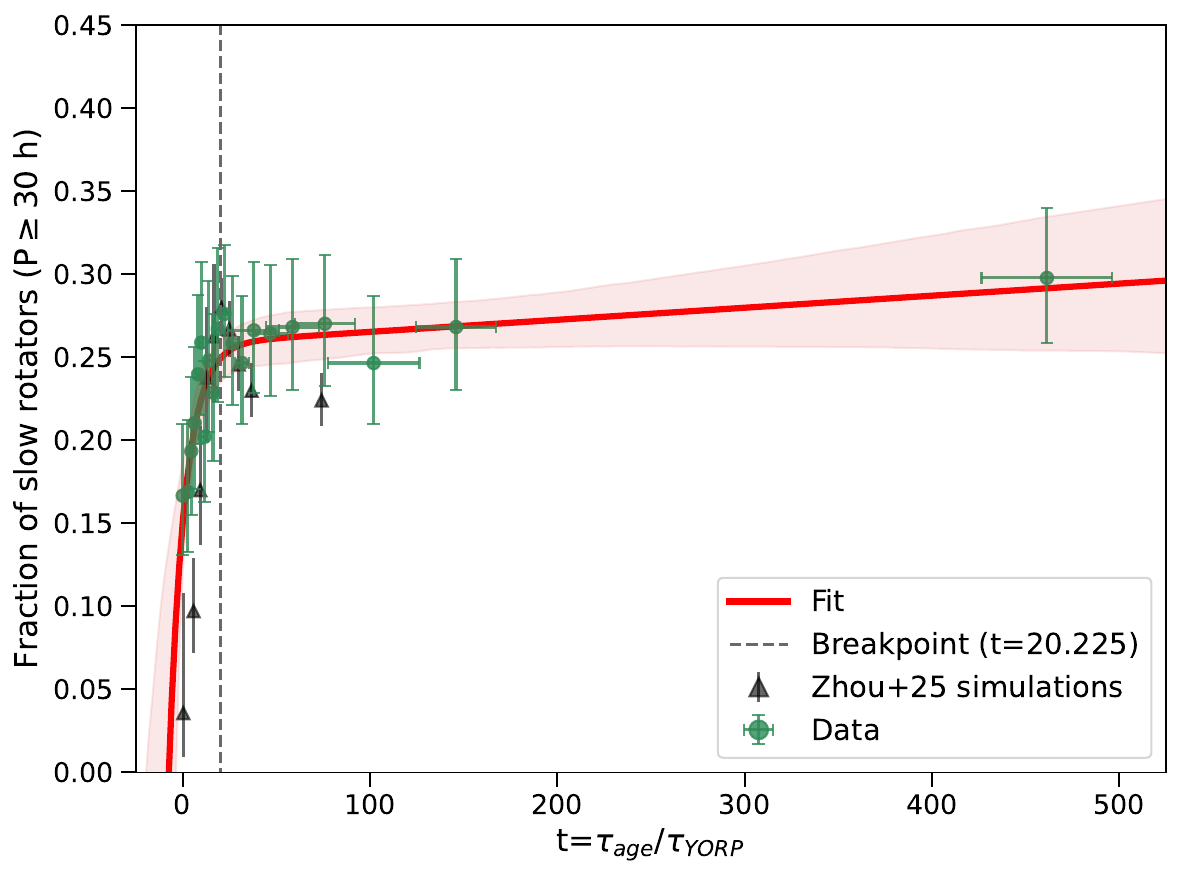}
    \caption{Fraction of slow rotators as a function of the dimensionless time $t$. {The green points are the binned collected observational data in this work. Black triangles are simulation data from \citet{wenhan_slowrot_25} model where the YORP timescale is manually increased by 10 times compared to the normal YORP timescale}. The red line is the fit function in Eq. \ref{eq:explin_fit}. The shaded area is the 95\% confidence interval of the fit. We do not show individual families because data are not sufficient for a statistical analysis.}
    \label{fig:slow_frac_fit}
\end{figure}

We binned the data with respect to $t$. For each time bin $t_i$, we calculated the fraction of slow rotators (i.e., asteroids with a rotation period $P\geq 30$ hours):
\begin{equation}
    f_{{\rm slow},i} = { N_i(P\geq{30\,\rm h}) \over N_i}\;.
\end{equation}
The error associated with the fraction is the 95\% confidence interval of a binomial distribution. The error model used to compute the uncertainties on the bins is described in Appendix \ref{sec:fit_details}.
By a visual inspection of the data, it was clear that there is an increasing trend up to a certain value of $t$ -- a 'breakpoint', after which the fraction of slow rotators remains roughly constant. For the analysis, we fixed the number of bins at 20: 10 bins before and 10 bins after the breakpoint. This is a good compromise between the number of bins and the number of objects in each bin.
The detailed fitting procedure is explained in Appendix \ref{sec:fit_details}.
The result for the best breakpoint is $t^*_{\text{bp}} = 20.225^{+1.063}_{-1.040}$. The final fit of the data was done using an exponential-linear function described as:
\begin{equation}
    f(t) = - \left(a\,e^{-jt} + mt +k\right) \;.
    \label{eq:explin_fit}
\end{equation}
The parameters for the best-fitted result are presented in Table \ref{tab:fit_results}.
Figure~\ref{fig:slow_frac_fit} shows the fitted function with the original asteroid data and synthetic data from simulation \citep{wenhan_slowrot_25} for comparison. The general trend of $f_{\rm slow}$ increases and asymptotically converges to $\sim 0.25$, a result that will be discussed in more detail in Section~\ref{sec:implications}.\\

{We note that the distribution of data points in $t$ is not uniform. In particular, Fig.~\ref{fig:t_distr} shows an exponential decay of objects in the interval $t \sim 200-450$, which results in a visible gap in the binned trends shown in Fig.~\ref{fig:slow_frac_fit}. This feature arises, in the first place from the fact that we used a quantile function to bin the data (see Appendix~\ref{sec:fit_details}), and, in second place, from the combined effect of the discrete age distribution of asteroid families, the strong dependence of $t$ on asteroid diameter ($t \propto D^{-2}$), and observational incompleteness for small objects ($\lesssim 6$ km), especially in older families.
}

\subsection{Polarization fraction}\label{sec:pol_frac}
Another interesting feature to study is the distribution of spin poles inside a V-shape. It is known that the Yarkovsky effect also depends on the obliquity of the spin vector \citep{Bottke2006}. The Yarkovsky effect is responsible for the V-shape distribution of an asteroid family in the (a-1/D) space \citep{michel_yarkovsky_2015}. Prograde rotators drift outward in semi-major axis, while retrograde rotators drift inward. We therefore classify an asteroid as aligned when its spin sign, relative to the orbital plane, matches the position according to the expected drift direction. {We also included spins with obliquities $\varepsilon \sim90^\circ$. We are aware that this may introduce some ambiguity, due to the errors on the spin axis direction (several degrees), that may result in similar probabilities for both spin directions. However, this concerns a small number of cases that do not change the global statistics: asteroids with $80^\circ < \varepsilon<100^\circ$ correspond to the 1.82\% of the dataset.}\\
The polarization fraction in each bin is then
\begin{equation}
    f_{{\rm pol}, i} = { N_i({\rm aligned \,\, objects}) \over N_i}\; .
\end{equation}
We again binned the data into 10 bins before and 10 bins after the breakpoint. 

\begin{figure}
    \centering
    \includegraphics[width=0.49\textwidth]{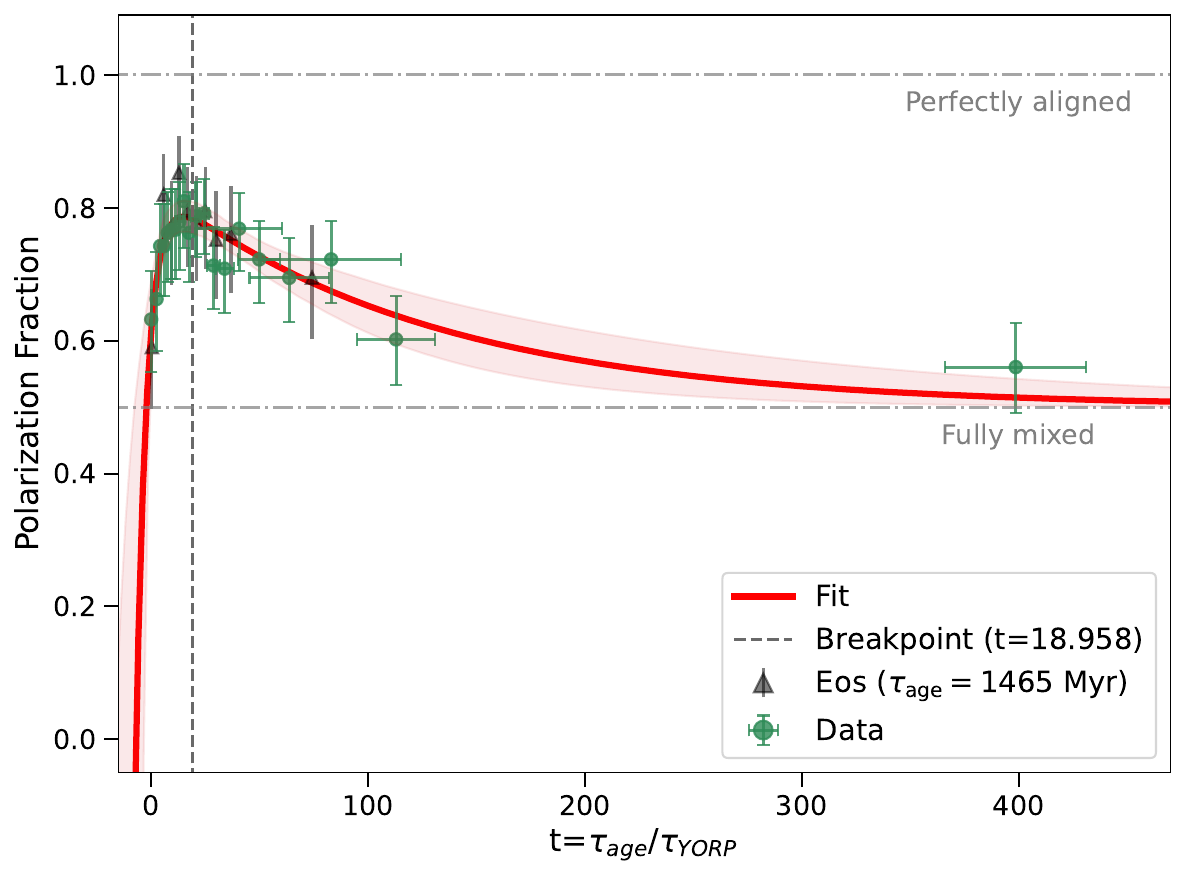}
    \caption{Polarization fraction as a function of the dimensionless time $t$. The green points are the binned data, the red line is the fit function in Eq. \ref{eq:polarization_fit}. The black triangles are the distribution of polarization fraction for the Eos family. The general trend is visible in single asteroid families. The shaded area is the 95\% confidence interval of the fit.}
    \label{fig:pol_frac_fit}
\end{figure}

The result for the best breakpoint is $t^*_{\text{bp}}=18.958_{-1.095}^{+1.228}$. The final values for the parameters were found fitting the following function to the data:
\begin{equation}
    f(t) = -a\, e^{-jt} + b\, e^{-gt} + 0.5 \; .
    \label{eq:polarization_fit}
\end{equation}
The second half of the equation is an exponential decay function that approaches the value of 0.5, which is the expected value of the polarization fraction for a random distribution of spin vectors.
The final result for the parameters is presented in Table \ref{tab:fit_results}.
The asteroid data and our fit are shown in Fig. \ref{fig:pol_frac_fit}, which shows that the function reaches the maximum value of $\simeq$ 80\% at $t\simeq15.07$.\\

{A similar gap in the $t \sim 150-400$ range is visible in the polarization fraction (Fig.~\ref{fig:pol_frac_fit}) and has the same origin discussed in Sect.~\ref{sec:slow_frac}.
}

\subsection{Assessing uncertainties}\label{sec:errors}
\begin{figure}
    \centering
    \includegraphics[width=0.48\textwidth]{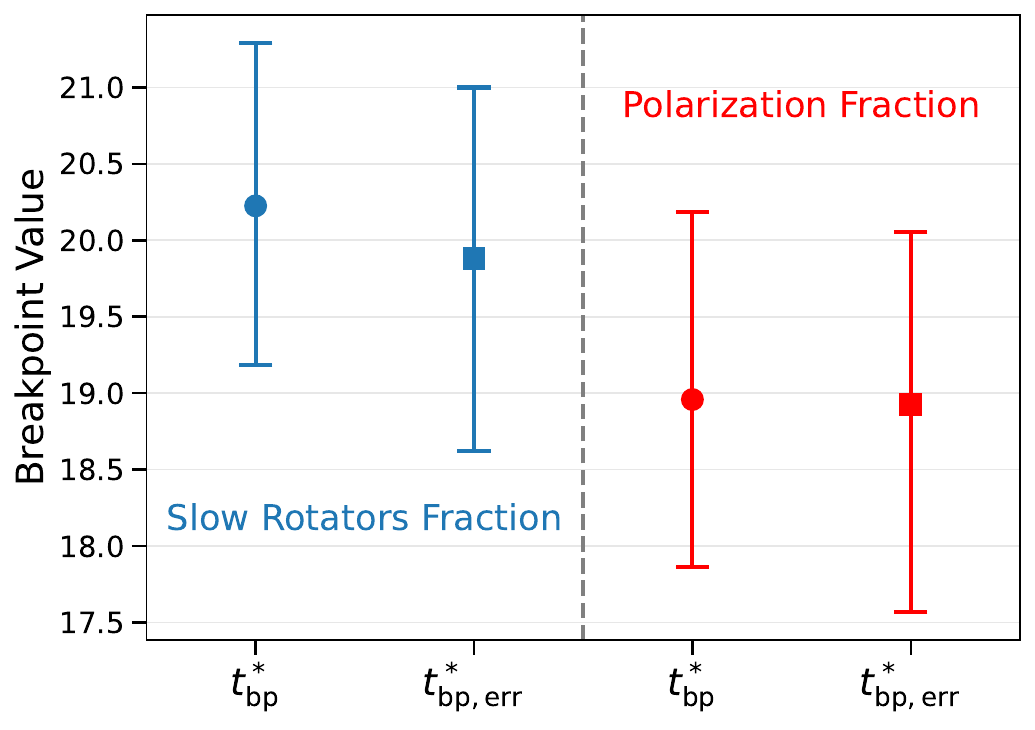}
    \caption{Comparison between the best breakpoint estimated from the original dataset, $t_{\rm bp}^*$ (circle marker) and the best breakpoint derived from the resampled dataset, $t_{\rm bp, err}^*$ (square marker). The breakpoints related to the slow rotators fraction (polarization fraction) are shown in blue (red). The breakpoints calculated from the resampled datasets are consistent with the original breakpoints.}
    \label{fig:bp_err}
\end{figure}

{Each asteroid in the dataset is associated with uncertainties in its diameter, $D_{\rm i,err}$, and in the estimated age of its parent family, $\tau_{\rm age,err}$. Accounting for these sources of error may lead to variations in the inferred value of the optimal breakpoint, $t_{\rm bp}^*$. In addition, the uncertainty on the family age reflects possible methodological limitations in the age determination and in the calibration of the Yarkovsky drift.\\
To assess the sensitivity of our results to these uncertainties, we performed a Monte Carlo resampling of the dataset. For each asteroid, we independently resampled the diameter and family age according to $D_{\rm i,new} \sim \mathcal{N}(D_{\rm i}, D_{\rm i,err})$ and $\tau_{\rm age,new} \sim \mathcal{N}(\tau_{\rm age}, \tau_{\rm age,err})$, and recomputed the corresponding dimensionless time $t$. The resampled dataset was then analyzed using the same fitting procedure described in the previous sections and detailed in Appendix~\ref{sec:fit_details}, yielding a new estimate of the best breakpoint. This procedure was repeated 500 times to build a distribution of breakpoint values, from which we computed the median and the
95\% confidence interval.\\
The resulting best estimations of the uncertainty-propagated breakpoints are shown in Fig.~\ref{fig:bp_err}, where they are compared with the original estimates. For the slow rotators fraction, we obtain $t_{\rm bp,err}^* = 19.883^{+1.116}_{-1.260}$, while for the polarization fraction we find $t_{\rm bp,err}^* = 18.926^{+1.129}_{-1.356}$.}\\

{To quantify the agreement between the original and uncertainty-propagated breakpoint estimates, we employed a two-tailed probability-to-exceed (P.T.E.) test \citep{raveri_2019}. We first computed the absolute difference between the median values, $|t_{\rm bp}^* - t_{\rm bp,err}^*|$, referred to as the `observed difference'. We then resampled with replacement both the original and the uncertainty-propagated breakpoint distributions and computed the difference between their medians for each realization. The P.T.E. was finally obtained as the fraction of resampled differences
exceeding the observed difference in absolute value.\\
The inclusion of uncertainties in diameter and family age produces only minor variations in the inferred breakpoint locations. The uncertainty-propagated estimates show substantial overlap with the original determinations, with largely overlapping 95\% confidence intervals and median values consistent within the corresponding uncertainty bands (Fig.~\ref{fig:bp_err}). The resulting P.T.E. values are $\simeq 48.4\%$ for the slow-rotator fraction and $\simeq 71.6\%$ for the polarization fraction, indicating statistically compatible results and supporting the robustness of the original breakpoint estimates.
}

\section{Implications}\label{sec:implications}

\subsection{Excess of slow rotators}
Our studies confirm the excess of slow rotators for the overall asteroid population. Significantly, for the first time, we capture the process of its occurrence.
We observe a rapid evolution in the fraction of slow rotators before the transition ($t < 20.5$), after which it converges to $\sim 25\%$. This fraction is consistent with a previous study by \citet{Pravec2008}. In their work, $f_{\rm slow} \sim 20\%$ is derived from 268 asteroids with diameters D=3-15 km. Of them, 79 are members of Hungaria/Phocaea families and most of them are older than 4 Gyr. 

In early studies, this excess was empirically modeled, either by artificially reducing spin acceleration by a factor of two \cite{Pravec2008} or by stopping rotational evolution \cite{bottke_search_2015} beyond a prescribed period. However, these approaches are not grounded in a self-consistent physical mechanism.

In \citet{wenhan_slowrot_25}'s model, tumbling is introduced for the first time and the YORP torque acting on tumblers is reduced by a factor $\eta_{\rm tumbling} = 10$, due to isotropic radiation under a chaotic tumbling rotation. The free parameter $\eta_{\rm tumbling}$ is calibrated to reproduce the slow-rotator fraction measured by \citet{Pravec2008}. Consequently, the tumblers spin up or down more slowly than spinners (i.e., those in principle-rotation), leading to an accumulation of asteroids in slow tumbling states and naturally producing an excess of slow rotators. Therefore, $f_{\rm slow}$ is directly tied to the fraction of tumblers and increases with both $\eta$ and the probability of a negative YORP torque. Figure~\ref{fig:slow_frac_fit} illustrates the evolution of $f_{\rm slow}$ in \citet{wenhan_slowrot_25}'s model, for $\eta_{\rm tumbling} = 50$.

\subsection{The stochastic YORP timescale}

We find that at $t \sim 20.23$, $f_{\rm slow}$ starts to converge. In both \citet{Pravec2008}'s and \citet{wenhan_slowrot_25}'s models, this corresponds to $\sim 2$ effective YORP timescales. The ``static'' YORP timescale $\tau_{\rm YORP}$ (Eq.~\ref{eq:yorp_time_scale}) refers to the time required for a body to evolve from its initial spin state to an end state under a fixed, unchanging YORP torque. In reality, however, YORP torques, and thus the evolution of the spin vector of an asteroid, often undergo a random walk \citep{bottke_search_2015}, because YORP is highly sensitive to small-scale surface features, such as craters and boulders that may evolve \citep{statler_extreme_2009, Golubov2012, cotto-figueroa_coupled_2015, Golubov2022, Zhou2022, Zhou2024}. 

To account for these stochastic processes, the effective YORP timescale is better represented by the ``stochastic YORP timescale'' rather than the ``static'' YORP timescale \citep{bottke_search_2015}, which can be estimated by
\begin{equation}
    \tau_{\rm YORP,stoc} = \eta_{\rm stoc} \cdot \tau_{\rm YORP}\; ,
\end{equation}
where $\eta_{\rm stoc}$ describes the random-walk evolution of the YORP torque over the body’s collisional history. This modifies not only the rotation evolution rate but also the Yarkovsky-driven spreading and V-shape evolution of asteroid families. However, $\eta_{\rm stoc}$ remains largely uncertain due to the complexity of the models.

This study provides the first direct observational validation of these theoretical predictions by examining the rotational evolution of asteroid families spanning a broad range of ages. {The rotational model predicts that $f_{\rm slow}$ converges at $\sim 2 \,\tau_{\rm YORP,stoc}$ \citep{Pravec2008,wenhan_slowrot_25}. \textbf{Fig.~\ref{fig:slow_frac_fit}} shows both the observational asteroid data collected in this study and the simulation data from \citet{wenhan_slowrot_25}'s model where the YORP timescale is manually increased by a factor of 10. The agreement between the simulation data and observations supports the stochastic YORP timescale is roughly an order of magnitude longer than the classical YORP timescale (i.e., $\eta_{\rm stoc} \sim 10$).}

Given such a long stochastic YORP timescale, in combination with the interplay of collisions, YORP self-limitation could be triggered. This phenomenon could explain why small asteroids evade transformation into top-shapes and/or binaries. This is achieved by constraining their rotational rates, thereby regulating the amount of angular momentum that can be imparted to a deformable body \citep{cotto-figueroa_coupled_2015}.
With YORP evolution proceeding $\sim 20$ times slower than classically predicted, many asteroids simply do not experience sufficient YORP cycles within their dynamical lifetimes to undergo catastrophic reshaping.

\subsection{Polarization fraction}
The evolution of the polarization fraction provides insight into the interplay between YORP-driven spin evolution and collisional reorientation. At early evolutionary stages ($t\lesssim 18.96$), the polarization fraction increases, reaching a maximum of $\simeq$ 80\% at $t\simeq15.07$. This rise indicates that Yarkovsky initially dominates the {orbital evolution, in the semi-major axis space}, systematically aligning asteroid spin axes with their orbital drift direction. In this initial stage, the YORP effect could increase the efficiency of the Yarkovsky drift by driving asteroids toward extreme obliquities (0$^\circ$ or 180$^\circ$), thereby enhancing the correlation between spin orientation and semi-major axis position within the V-shape.
However, the fact that polarization reaches only 80\% rather than complete alignment suggests a combination of stochastic re-orientation due to collisions and errors in the spin currently determined spin directions, the precise proportions of which remain unknown.\\
The subsequent decay of polarization to 0.5 at larger $t$ indicates complete randomization of spin axes relative to orbital position. This transition suggests that collisional reorientation becomes increasingly important relative to Yarkovsky- and YORP-driven evolution at extended timescales. The cumulative effect of non-catastrophic collisions progressively breaks the connection between the current spin state and the Yarkovsky drift history that determined the asteroid's position in the V-shape. This interpretation is consistent with theoretical models suggesting that for smaller objects (i.e., more evolved family members), collisions occur more frequently and with relatively greater strength, causing bodies to transition between different spin states more rapidly \citep{marzari_combined_2011}.  

The randomization of spin axes through collisions has implications for the interpretation of asteroid family ages derived from the V-shape. The fundamental hypothesis underlying V-shape age dating is that the direction and drift rate of the semi-major axis {for family members located at the edges of the V-shape,} remain constant over time \citep{MILANI201446}. However, collisional spin reorientation introduces a random walk component to the Yarkovsky drift. Under these conditions, the inverse slopes of the V-shape borders are no longer proportional to the family age, but rather scale with the square root of the family age \citep{marzari_combined_2011}. The decay of polarization toward a random state at large dimensionless times indicates that collisional spin reorientation becomes increasingly important for family members older than $t \approx 20$, marking the limit beyond which V-shape age estimates become unreliable. For older families, partial erasure of the Yarkovsky signature may bias V-shape results. In such cases, the degree of polarization could serve as an independent diagnostic or correction factor. Further work is needed to quantify these effects.

\section{Conclusions}\label{sec:conclusions}

In this work, we utilized the rapidly expanding dataset of asteroid spin states to empirically investigate the long-term rotational evolution of main-belt asteroid families. By introducing a dimensionless evolutionary timescale ($t = \tau_{\rm age}/\tau_{\rm YORP}$), we directly compared 28 asteroid families with a wide range of ages and provided new observational constraints on the coupled YORP-Yarkovsky collision processes.

We showed that the fraction of slow rotators increases rapidly with $t$ and saturates at $\sim 0.25$ near $t \approx 20$. This behavior matches theoretical predictions for the onset of stochastic YORP self-limitation and yields an effective stochastic YORP timescale 10 times longer than the classical formulation ($\eta_{\rm stoc} \sim 10$). Our results therefore provide the first direct population-level confirmation that YORP evolution is significantly slowed by surface-morphology–driven torque variability.

We further showed that the polarization fraction -- quantifying the spin–orbit correlation within V-shapes -- initially rises to $\sim 80\%$ as YORP drives bodies toward extreme obliquities and efficient Yarkovsky drift. However, beyond $t \approx 20$, the polarization decays toward the random limit $\sim 50\%$, indicating that collisional reorientation increasingly dominates the spin evolution of older families. This transition marks the approximate age boundary beyond which V-shape estimates become systematically biased, and suggests that the degree of polarization may be used as a diagnostic or correction factor for advanced families.

The rotational evolution trends of $f_{\rm slow}$ and $f_{\rm pol}$ can be further tested with forthcoming LSST observations, which are expected to increase the available spin-state sample by at least an order of magnitude \citep{LSST}. With such datasets, these trends could provide an additional dimension for constraining asteroid family ages, by fitting $\tau_{\rm age}$ to the evolutionary tracks in Figs.~\ref{fig:slow_frac_fit} and \ref{fig:pol_frac_fit} for individual families.

\begin{acknowledgements}
W.-H. Zhou acknowledges funding support from the Japan Society for the Promotion of Science (No. P25021). G. Bertinelli thanks G. Viterbo for helpful discussions on topics related to this work. The authors thank the anonymous referee for productive comments and suggestions.
\end{acknowledgements}

\bibliographystyle{aa}
\bibliography{biblio}{}

\begin{appendix}
\onecolumn



\section{Asteroid families}

\begin{table}[htbp]
  \centering
  \caption{{List of the 28 families used in the analysis. Ages and associated errors were calculated by \citet{spoto_asteroid_2015}. From the intersection of the two sides of the V-shape, for each family, we calculated $a_{\rm center}$. We list also the number of objects with spin period and cosine of the obliquity determination.}}
    \begin{tabular}{cccccc}
    \hline
    \hline
    Family & Age IN/OUT [Myr] & err\_age IN/OUT [Myr] & $a_{\rm center}$ [AU] & \# periods & \# obliquities \\
    \hline
    3 Juno & 550/370 & 156/161 & 2.676 & 66    & 15 \\
    4 Vesta & 930/1906 & 217/659 & 2.343 & 1086  & 378 \\
    10 Hygiea & 1330/1368 & 300/329 & 3.163 & 342   & 171 \\
    15 Eunomia & 1955/1144 & 421/236 & 2.646 & 1822  & 711 \\
    20 Massalia & 174/189 & 35/41 & 2.409 & 156   & 30 \\
    24 Themis & 2447/3782 & 836/958 & 3.141 & 814   & 355 \\
    31 Euphrosyne & 1309/1160 & 312/272 & 3.157 & 134   & 67 \\
    158 Koronis & 1792/1708 & 444/399 & 2.886 & 674   & 349 \\
    163 Erigone & 212/230 & 68/50 & 2.373 & 38    & 5 \\
    221 Eos & 1412/1537 & 290/334 & 3.022 & 2252  & 1198 \\
    283 Emma & 290/628 & 67/234 & 3.047 & 47    & 28 \\
    302 Clarissa & 57/47 & 18/12 & 2.404 & 4     & 2 \\
    396 Aeolia & 100/91 & 31/27 & 2.745 & 8     & 5 \\
    434 Hungaria & 208/205 & 65/62 & 1.936 & 216   & 36 \\
    480 Hansa & 763/950 & 346/223 & 2.624 & 167   & 78 \\
    569 Misa & 319/249 & 255/105 & 2.665 & 25    & 7 \\
    606 Brangane & 48/44 & 11-Oct & 2.583 & 7     & 3 \\
    668 Dora & 532/471 & 159/184 & 2.785 & 235   & 92 \\
    808 Merxia & 338/321 & 73/69 & 2.746 & 67    & 26 \\
    845 Naema & 149/163 & 35/34 & 2.938 & 28    & 13 \\
    847 Agnia & 1003/669 & 288/167 & 2.782 & 145   & 60 \\
    1040 Klumpkea & 664/661 & 221/244 & 3.124 & 213   & 82 \\
    1128 Astrid & 150/150 & 32/32 & 2.788 & 16    & 7 \\
    1303 Luthera & 279/273 & 88/89 & 3.218 & 29    & 16 \\
    1547 Nele & 14/15 & 5/7   & 2.644 & 11    & 1 \\
    1726 Hoffmeister & 337/328 & 96/92 & 2.786 & 65    & 29 \\
    3330 Gantrisch & 436/492 & 168/197 & 3.149 & 54    & 23 \\
    3815 Konig & 51/51 & 14/14 & 2.575 & 18    & 7 \\
    \hline
    \end{tabular}%
  \label{tab:families}%
\end{table}%

\section{Fitting procedure}\label{sec:fit_details}
Since the distribution of $t$ (Eq. \ref{eq:t}) is not uniform (see Fig.~\ref{fig:t_distr}), we used the quantile-based discretization function\footnote{\href{https://pandas.pydata.org/docs/reference/api/pandas.qcut.html}{\texttt{pandas.qcut}}} to roughly have the same number of asteroids in each bin. The value of $t$ at which each bin is centered is associated with an error, given by the standard deviation of the dimensionless times inside the bin. The fraction of slow rotators in each bin is calculated as the number of slow rotators divided by the total number of asteroids in the bin. The error associated with the fraction is the 95\% confidence interval of a binomial distribution. In particular, we used the Wilson score interval with continuity correction \citep{wilson_score}.\\

We fitted the data with respect to the slow rotators fraction (Sect.~\ref{sec:slow_frac}), with a piecewise function defined as:
\begin{equation}
f(t) = 
\begin{cases}
    - \left(a e^{-jt} + k\right) & \text{if } t \leq t_{\text{bp}} \\
    - \left(m t + b\right) & \text{if } t > t_{\text{bp}} \; ,
\end{cases}
\label{eq:piecewise_fit}
\end{equation}
where $t_{\text{bp}}$ is the breakpoint and $a$, $j$, $k$, $m$, and $b$ are the parameters to be fitted. 

\textbf{For parameter estimation}, we employed a weighted Huber loss function to ensure robustness against outliers:

\begin{equation}
    L(r) = 
    \begin{cases}
        \frac{1}{2}r^2 & \text{if } |r| \leq \delta \\
        \delta \left(|r| - \frac{1}{2}\delta\right) & \text{otherwise}
    \end{cases}
    \quad \text{with}\;r=\frac{y-f(t)}{\sqrt{\sigma^2}}\;, 
    \label{eq:huber_loss}
\end{equation}
where $\delta=1.345$ \citep{huber_robust_1964}. The weighted residual $r$ accounts for error propagation through $\sqrt{\sigma^2}=\sqrt{y_{\text{err}}^2+(\partial_tf(t)\cdot t_{\text{err}})^2}$, where $y_{\text{err}}$ and $t_{\text{err}}$ are the errors associated with the fraction of slow rotators (polarization) and with $t$, respectively. Since the error on the fraction of slow rotators (polarization) is asymmetric (from the Wilson confidence interval), we adopted the following scheme:
\begin{equation*}
    y_{\text{err}} = 
    \begin{cases}
        y_{\text{upper}} - y & \text{if } r \geq 0 \\
        y - y_{\text{lower}} & \text{otherwise}\;.
    \end{cases}
\end{equation*}
The total loss was calculated as $\mathcal{L}_{\text{Huber}} = \sum_{i} L(r_i)$. The Huber loss provides robust parameter estimates that are less sensitive to outliers compared to ordinary least squares, and does not require errors to be normally distributed \citep{huber_robust_1964}.

\textbf{For breakpoint selection}, {we implemented an iterative scan over candidate $t_{\text{bp}}$ values to identify the optimal model complexity. At each candidate breakpoint, we fitted the model parameters using the Huber loss, then evaluated model quality using the Akaike Information Criterion (AIC):
\begin{equation}
    \text{AIC} = 2k + n\ln(\text{RSS}/n)\,\, ,
\end{equation}
where $k$ is the effective number of free parameters, $n$ is the number of data points, and RSS is the weighted Residual Sum of Squares:
\begin{equation}
    \text{RSS} = \sum_i \left(\frac{y_i - f(t_i)}{\sqrt{\sigma_i^2}}\right)^2\;.
\end{equation}
Note that while parameter estimation used the Huber loss, model selection employed RSS for AIC calculation to maintain consistency with standard information-theoretic model comparison \citep{burnham_model_2004}. The AIC penalizes model complexity and prevents overfitting, with lower values indicating better models. The breakpoint that minimized AIC was selected as optimal.}

The parameters were fitted using the Sequential Least Squares Programming (SLSQP) algorithm from the \href{https://docs.scipy.org/doc/scipy/index.html}{\texttt{SciPy}} package in Python, with an equality constraint enforcing the continuity of $f(t)$ at $t_{\text{bp}}$.\\

The scan was initialized with a wide range of $t_{\text{bp}}$ values and then refined by manually reducing, in an iterative fashion, the interval around the values that provided the lowest AIC value. 
To have a statistical estimate of the parameters, we performed a Monte Carlo bootstrap sampling of the data, i.e., we randomly sampled the data with replacement and fitted the piecewise function to each set. We repeated the process 500 times. The values of the final parameters are the median of the distribution obtained from the procedure, while the errors are given by the 95\% confidence interval of the distributions.\\

The fitting procedure for the polarization fraction (Sect. \ref{sec:pol_frac}) is the same as explained above, and the best breakpoint $t^*_{\text{bp}}$ was found by fitting the following function:
\begin{equation}
    f(t) = 
    \begin{cases}
        - \left(a\, e^{-jt} + k\right) & \text{if } t \leq t_{\text{bp}} \\
        b\, e^{-gt} + 0.5 & \text{if } t > t_{\text{bp}} \; .
    \end{cases}
    \label{eq:pol_frac_fit}
\end{equation}
The second part of the piecewise function is an exponential decay function that approaches the value of 0.5, which is the expected value of the polarization fraction for a random distribution of spin vectors.\\

In the end, for both the slow rotators fraction and the polarization fraction, we fitted the data with a unique function (Eq. \ref{eq:explin_fit} and Eq. \ref{eq:polarization_fit}, respectively), fixing the breakpoint at $t^*_{\text{bp}}$. The final parameters were again obtained from a Monte Carlo bootstrap sampling after 1000 simulations.

\renewcommand{\arraystretch}{1.5} 
\begin{table}[!h]    
    \centering
    \caption{Parameters of the fit functions from Eq. \ref{eq:explin_fit} and Eq. \ref{eq:polarization_fit}. The parameter value is the median of the distribution, and the error is given by the 95\% confidence interval.}
    \label{tab:fit_results}
    \begin{tabular}{cc||cc}
    \hline
    \hline
    Parameter Eq. \ref{eq:explin_fit}        & Value                                           & Parameter Eq. \ref{eq:polarization_fit}        & Value                       \\ 
    \hline
    \multicolumn{1}{c}{a} & \multicolumn{1}{c||}{$0.109_{-0.038}^{+0.037}$}  & \multicolumn{1}{c}{a} & $0.226_{-0.085}^{+0.102}$   \\
    \multicolumn{1}{c}{j} & \multicolumn{1}{c||}{$0.117_{-0.061}^{+0.099}$}  & \multicolumn{1}{c}{j} & $0.193_{-0.106}^{+0.203}$  \\
    \multicolumn{1}{c}{m} & \multicolumn{1}{c||}{${-7.30\text{e-}5}_{-1.17\text{e-}4}^{+1.05\text{e-}4}$} & \multicolumn{1}{c}{b} & $0.337_{-0.043}^{+0.108}$   \\
    \multicolumn{1}{c}{k} & \multicolumn{1}{c||}{$-0.258_{-0.021}^{+0.016}$} & \multicolumn{1}{c}{g} & $0.008_{-0.003}^{+0.005}$    \\
    \hline
    \end{tabular}
\end{table}
\renewcommand{\arraystretch}{1.0} 

\end{appendix}

\end{document}